\documentclass[twocolumn]{autart}

\usepackage[dvips]{graphicx,color}
\usepackage[dvips]{epsfig}
\usepackage{mathptmx}
\usepackage{amsmath}
\usepackage{amssymb}
\usepackage{xcolor}
\usepackage{color}
\usepackage{array}
\usepackage{graphicx}
\usepackage{subfig}
\usepackage{savesym}
\savesymbol{AND}
\usepackage{amssymb}
\usepackage{amsmath}
\usepackage{amsfonts}
\usepackage{mathrsfs}
\usepackage{cases}
\usepackage{epstopdf}
\usepackage{leftidx}
\usepackage{extarrows}
\usepackage{colortbl}
\usepackage{multirow}
\usepackage{booktabs}
\usepackage{amssymb}  
\usepackage{algorithm}
\usepackage{algorithmic}
\usepackage{multirow}
\usepackage{tikz}
\usepackage{comment}

\theoremstyle{definition}
\newtheorem{rmk}{Remark}
\newtheorem{lemm}{Lemma}
\newtheorem{thme}{Theorem}
\newtheorem{dfn}{Definition}
\newtheorem{prop}{Proposition}
\newtheorem{ass}{Assumption}
\newtheorem{cllry}{Corollary}

\renewcommand{\qed}{\hfill\blacksquare}

\allowdisplaybreaks

\begin{document}

\begin{frontmatter}

\title{Sampled-Data and Event-triggered Boundary Control of a Class of Reaction-Diffusion PDEs with Collocated Sensing and Actuation}

\thanks[footnoteinfo]{Corresponding author B.~Rathnayake Tel. +1(518)5965193.}

\author[Paestum]{Bhathiya Rathnayake}\ead{rathnb@rpi.edu},    
\author[troy]{Mamadou Diagne}\ead{diagnm@rpi.edu},               
\author[Baiae]{Iasson Karafyllis}\ead{ iasonkar@central.ntua.gr}  

\address[Paestum]{Department of Electrical, Computer, and Systems Engineering, Rensselaer Polytechnic Institute, New York, USA}  
\address[troy]{Department of Mechanical, Aerospace, and Nuclear Engineering, Rensselaer Polytechnic Institute, New York, USA}
\address[Baiae]{Department of Mathematics, National Technical University of Athens, Greece}        

\begin{abstract}
This paper provides observer-based sampled-data and event-triggered boundary control strategies for a class of reaction-diffusion PDEs with collocated sensing and Robin actuation. Infinite-dimensional backstepping design is used as the underlying control approach. It is shown that the continuous-time output feedback boundary control applied in a sample-and-hold fashion ensures global closed-loop exponential stability, provided that the sampling period is sufficiently small. Further, robustness to perturbations of the sampling schedule is guaranteed. For the event-triggered implementation of the continuous-time controller, a dynamic triggering condition is utilized. The triggering condition determines the time instants at which the control input needs to be updated. Under the observer-based event-triggered boundary control, it is shown that there is a minimal dwell-time between two triggering instants independent of initial conditions. Further, the global exponential convergence of the closed-loop system to the equilibrium point is established.  A simulation example is provided to validate the theoretical results.

\end{abstract}

\end{frontmatter}

\section{Introduction}
Computer-controlled systems often consist of continuous-time plants and digital computers that interact via a feedback channel to achieve specific control objectives. For such systems, periodic sampling/update of control inputs is sometimes not desirable due to limitations in the hardware, software, and communication resources \cite{hespanha2007survey}. Therefore, control strategies that rely on non-uniform sampling schedules, known as aperiodic sampled-data control, have been introduced \cite{fridman2004robust,karafyllis2009global,nesic2009explicit,karafyllis2011nonlinear,karafyllis2012global,pepe2016stability}. A major drawback in most of the existing aperiodic sampled-data control methods is the lack of explicit criteria for selecting appropriate sampling schedules required for the controller implementation. This has led to the development of event-triggered control strategies \cite{tabuada2007event,heemels2012introduction,6069816,heemels2012periodic,marchand2012general,tallapragada2013event,postoyan2014framework,girard2014dynamic}, where the instances of control updates are determined by occurrences of events that indicate the need for fresh control updates. Thus, event-triggered control provides a rigorous resource-aware method of implementing control laws into digital platforms. 

Sampled-data control of infinite-dimensional systems has been studied in works such as \cite{logemann2003stability,logemann2005generalized,rebarber2006robustness,fridman2012robust,selivanov2017sampled,karafyllis2017sampled,karafyllis2018sampled,kang2018distributed,wang2019mixed}. The major works \cite{logemann2003stability,rebarber2006robustness} provide necessary and sufficient conditions for periodic sampled-data control of general infinite-dimensional systems, which have been extended in \cite{logemann2005generalized} to incorporate generalized sampling. Sampled-data controllers for parabolic PDEs using matrix inequalities have been proposed in \cite{fridman2012robust,selivanov2017sampled,kang2018distributed}. Using small-gain arguments, the authors of \cite{karafyllis2018sampled} obtain results that guarantee closed-loop exponential stability for 1-D parabolic PDEs under Zero-Oder-Hold implementations of continuous-time boundary feedback designs (emulation). Similar results have been obtained for 1-D linear transport PDEs with non-local terms in \cite{karafyllis2017sampled}. A distributed sampled-data control approach using a finite number of local piecewise measurements in space is proposed for semilinear parabolic PDEs in \cite{wang2019mixed}. Via modal decomposition, the authors of \cite{katz2021delayed} present sampled-data control approaches under boundary or distributed sensing together with non-local actuation, or Dirichlet actuation with distributed sensing for reaction-diffusion systems.

Several recent works that employ event-triggered boundary control strategies on infinite-dimensional systems can be found in \cite{selivanov2016distributed,espitia2016event,espitia2017event,wang2019observer,espitia2020observer,espitia2021event,diagne2020event,katz2020boundary}.  In general, event-triggered control includes two main components: a feedback control law that stabilizes the system and an event-triggered mechanism that determines when the control value has to be updated.  All event-triggered control designs should be free of the Zeno phenomenon \cite{heemels2012introduction}; otherwise, the design would be infeasible for digital implementation due to the triggering of an infinite number of control updates over a finite period. Zeno-free behavior is often ensured by showing a guaranteed lower bound for the time between two adjacent events, known as minimal dwell-time. 

Both Sampled-data and event-triggered boundary control of linear parabolic PDEs with boundary sensing only are pretty challenging. The actuation type and the sensor configuration have to be carefully selected to obtain successful control designs. A general robustness result that guarantees closed-loop exponential stability under the emulation of continuous-time boundary output feedback control designs with a sufficiently small sampling period (not necessarily periodic) is missing for sampled-data control with boundary sensing and actuation. Sampled-data backstepping with Neumann actuation leads to an undesirable trace term that arises from the difference between the continuous-time feedback and the applied control action (the input holding error), for which it is impossible to obtain a useful bound on its rate of convergence. Furthermore, sampled-data backstepping boundary output feedback requires a bound for the local output making Dirichlet type sensing is the only viable option. The Dirichlet output can be bounded by using the $H^{1}$-norm estimate of the observer error target system. However, such $H^1$-norm estimates only exist when the boundary condition at the uncontrolled end is Dirichlet (irrespective of the boundary conditions at the controlled end) \cite{karafyllis2019input}. Therefore, boundary sensing anti-collocated with the actuation is not conducive for sampled-data backstepping control.

Under event-triggered boundary control of general parabolic PDEs with boundary observation, the possibility of avoiding the Zeno phenomenon is unknown. The actuation type is critical as both Dirichlet and Neumann actuation under backstepping pose a severe impediment in proving the existence of a minimal dwell-time and hence well-posedness and convergence results due to unbounded local terms. Using the modal decomposition approach, the authors \cite{katz2020boundary} present an observer-based event-triggered boundary control of a reaction-diffusion equation with anti-collocated sensing and Robin boundary actuation in the presence of input and output time-varying delay. In \cite{rathnayake2020observer}, an event-triggered backstepping boundary control strategy for reaction-diffusion PDEs with anti-collocated sensing and Robin boundary actuation is proposed. \textcolor{red}{The authors of \cite{chen2017backstepping} propose a continuous-time backstepping-based boundary feedback control for a fractional reaction diffusion system with Robin boundary conditions.}

This paper considers Robin boundary control of reaction-diffusion PDEs with collocated sensing and actuation using the infinite-dimensional backstepping approach. We prove that there is a sufficiently small sampling period such that the global exponential stability of the closed-loop system is preserved under the emulation of the continuous-time observer-based controller. We also derive conditions for the (conservative) upper bounds of the sampling period and establish robustness to perturbations of the sampling schedule. Moreover, for the controller's practical implementation, we propose an event-triggered control strategy as in \cite{rathnayake2020observer} using a dynamic event-triggering condition under which we show that the Zeno phenomenon cannot occur. We prove the global exponential convergence of the closed-loop system subject to the proposed event-triggered control.

The paper is organized as follows. Section 2 introduces the class  of  linear  reaction-diffusion  system  and  the  continuous-time  output  feedback  boundary  control.  Section  3  presents the observer-based sampled-data boundary control.  In  Section  4,  we introduce the observer-based event-triggered boundary control.  We  provide  a  numerical  example  in  Section  5  to illustrate the results and conclude the paper in Section 6.

\textit{Notation:} $\mathbb{R}_{+}$ is the nonnegative real line whereas $\mathbb{N}$ is the set of natural numbers including zero.  By $C^{0}(A;\Omega)$, we denote the class of continuous functions on $A\subseteq\mathbb{R}^{n}$, which takes values in $\Omega\subseteq\mathbb{R}$. By $C^{k}(A;\Omega)$, where $k\geq 1$, we denote the class of continuous functions on $A$, which takes values in $\Omega$ and has continuous derivatives of order $k$.  $L^{2}(0,1)$ denotes the equivalence class of Lebesgue measurable functions $f:[0,1]\rightarrow\mathbb{R}$ such that $\Vert f\Vert=\big(\int_{0}^{1}f^2(x)dx\big)^{1/2}<\infty$. $H^{1}(0,1)$ denotes the equivalence class of Lebesgue measurable functions $f:[0,1]\rightarrow\mathbb{R}$ such that $\int_{0}^{1} f^2(x)dx+\int_{0}^{1} f_x^2(x)dx<\infty$. $H^{2}(0,1)$ denotes the Sobolev space of continuously differentiable functions on $[0,1]$ with measurable, square integrable second derivative. Let $u:[0,1]\times\mathbb{R}_{+}\rightarrow\mathbb{R}$ be given. $u[t]$ denotes the profile of $u$ at certain $t\geq 0$, \textit{i.e.,} $\big(u[t]\big)(x)=u(x,t),$ for all $x\in [0,1]$. For an interval $I\subseteq\mathbb{R}_{+},$ the space $C^{0}\big(I;L^{2}(0,1)\big)$ is the space of continuous mappings $I\ni t\rightarrow u[t]\in L^{2}(0,1)$. $I_{m}(\cdot), $ and $J_{m} (\cdot)$ with $m$ being an integer respectively denote the modified Bessel and (nonmodified) Bessel functions of the first kind.   
\section{Observer-based backstepping boundary control and emulation}
Let us consider the following 1-D reaction-diffusion system with constant coefficients:
\begin{subequations}\label{ctp}
\begin{align}\label{ctpe1}
u_{t}(x,t)&=\varepsilon u_{xx}(x,t)+\lambda u(x,t),\\
\label{ctpe2}
u(0,t)&=0,\\\label{ctpe3}
u_{x}(1,t)+qu(1,t)&=U(t),
\end{align}
\end{subequations}
and the initial condition $u[0]\in H^{1}(0,1),$ where $\varepsilon,\lambda>0,$ $u: [0,1]\times [0,\infty)\rightarrow\mathbb{R}$ is the system state, and $U(t)$ is the control input.
 \begin{ass}\label{ass1} 
The plant's  parameters $q$, $\lambda$, and $\varepsilon$ satisfy the following inequality:  $$q>\lambda/2\varepsilon.$$
\end{ass}
\begin{rmk}
\textcolor{red}{Assumption \ref{ass1} is required to avoid a trace term for which it is impossible to obtain a useful bound on its rate of change. In order to overcome this, it is required that $q-\lambda/2\varepsilon>0$. Furthermore, It should be mentioned that an eigenfunction expansion of the solution of \eqref{ctp} with $U(t)=0$ (zero input) shows that the system is unstable when $\lambda>\varepsilon\pi^2$, no matter what $q>0$ is.}
\end{rmk}

We propose an observer for the system \eqref{ctp} using $u(1,t)$ as the available measurement/output. Note that the output is collocated with the input. The observer consists of a copy of the system \eqref{ctp} with output injection terms, which is stated as follows:
\begin{subequations}\label{cto}\begin{align}\label{ctoe1}
\begin{split}
\hat{u}_{t}(x,t)&=\varepsilon \hat{u}_{xx}(x,t)+\lambda \hat{u}(x,t)\\&\qquad+p_1(x)\big(u(1,t)-\hat{u}(1,t)\big),
\end{split}
\\\label{ctoe2}
\hat{u}(0,t)&=0,
\\\label{ctoe3}
\hat{u}_{x}(1,t)+q\hat{u}(1,t)&=U(t)+p_{10}\big(u(1,t)-\hat{u}(1,t)\big),
\end{align}
\end{subequations}
and the initial condition $\hat{u}[0]\in H^{1}(0,1)$. Here, the function $p_{1}(x)$ and the constant $p_{10}$ are observer gains to be determined. 
\begin{rmk}
The restriction that the initial conditions should satisfy $u[0],\hat{u}[0]\in H^{1}(0,1)$ is only required for sampled-data control. For event-triggered control, it is sufficient that $u[0],\hat{u}[0]\in L^{2}(0,1)$.
\end{rmk}

Let us denote the observer error by $\tilde{u}(x,t)$, which is defined as\begin{equation}\label{diff}
\tilde{u}(x,t):=u(x,t)-\hat{u}(x,t).
\end{equation}By subtracting \eqref{cto} from \eqref{ctp}, one can see that $\tilde{u}(x,t)$ satisfies the following PDE:
\begin{subequations}\label{ctoe}
\begin{align}\label{ctoee1}
\begin{split}
\tilde{u}_{t}(x,t)&=\varepsilon\tilde{u}_{xx}(x,t)+\lambda\tilde{u}(x,t)-p_{1}(x)\tilde{u}(1,t),
\end{split}
\\\label{ctoee2}
\tilde{u}(0,t)&=0,
\\\label{ctoee3}
\tilde{u}_{x}(1,t)+q\tilde{u}(1,t)&=-p_{10}\tilde{u}(1,t).
\end{align}
\end{subequations}

\begin{prop}\label{prop1} 
Under the invertible backstepping transformation\begin{equation}\label{ctobt}
\tilde{u}(x,t)=\tilde{w}(x,t)-\int_{x}^{1}P(x,y)\tilde{w}(y,t)dy,
\end{equation}where \begin{equation}\label{solP}
P(x,y)=-\frac{\lambda}{\varepsilon} x\frac{I_{1}\big(\sqrt{\lambda(y^{2}-x^{2})/\varepsilon}\big)}{\sqrt{\lambda(y^{2}-x^{2})/\varepsilon}},
\end{equation}for $0\leq x\leq y\leq 1$, the observer error system \eqref{ctoe} with the gains $p_{1}(x)$ and $p_{0}$ chosen as\begin{align}\label{p1}
p_{1}(x)=-\varepsilon qP(x,1)-\varepsilon P_{y}(x,1),\hspace{10pt}p_{10}=-P(1,1)=\frac{\lambda}{2\varepsilon},
\end{align}
gets transformed to the following observer error target system
\begin{subequations}\label{ctots}
\begin{align}\label{ctotse1}
\tilde{w}_{t}(x,t)&=\varepsilon \tilde{w}_{xx}(x,t),
\\\label{ctotse2}
\tilde{w}(0,t)&=0,
\\\label{ctotse3}
\tilde{w}_{x}(1,t)&=-q\tilde{w}(1,t),
\end{align}
\end{subequations}
which is globally $L^{2}$-exponentially stable for any $q>0$. 
\end{prop}

\textit{Proof:} The proof is very similar to that of Proposition 1 in \cite{rathnayake2020observer} and hence omitted. \hfill $\qed$

The inverse transformation of \eqref{ctobt} can be shown to be as follows:
\begin{equation}\label{ibtoe}
\tilde{w}(x,t)=\tilde{u}(x,t)+\int_{x}^{1}Q(x,y)\tilde{u}(y,t)dy,
\end{equation}where $Q(x,y)$ is 
\begin{equation}\label{solQ}
Q(x,y)=-\frac{\lambda}{\varepsilon} x\frac{J_{1}\big(\sqrt{\lambda(y^{2}-x^{2})/\varepsilon}\big)}{\sqrt{\lambda(y^{2}-x^{2})/\varepsilon}},
\end{equation}
for $0\leq x\leq y\leq 1$.

\begin{prop}\label{prop2} The invertible backstepping transformation
\begin{equation}\label{ctbt}
\hat{w}(x,t)=\hat{u}(x,t)-\int_{0}^{x}K(x,y)\hat{u}(y,t)dy,
\end{equation}
where\begin{equation}\label{ctcks}
K(x,y)=-\frac{\lambda}{\varepsilon}y\frac{I_{1}\big(\sqrt{\lambda(x^{2}-y^{2})/\varepsilon}\big)}{\sqrt{\lambda(x^{2}-y^{2})/\varepsilon}},
\end{equation} for $0\leq y\leq x\leq 1,$  and a control law $U(t)$ chosen as\begin{equation}\label{ctcl}
U(t)=\int_{0}^{1}\Big(rK(1,y)+K_{x}(1,y)\Big)\hat{u}(y,t)dy,
\end{equation}map the system \eqref{cto} with the gains $p_{1}(x)$ and $p_{10}$ chosen as in \eqref{p1}, into the following target system:
\begin{subequations}\label{etots}\begin{align}\label{etotse1}
\hat{w}_{t}(x,t)&=\varepsilon \hat{w}_{xx}(x,t)+g(x)\tilde{w}(1,t),
\\\label{etotse2}
\hat{w}(0,t)&=0,
\\\label{etotse3}
\hat{w}_{x}(1,t)&=-r\hat{w}(1,t)+\frac{\lambda}{2\varepsilon}\tilde{w}(1,t).
\end{align}\end{subequations}
with\begin{equation}\label{gt}
g(x)=p_{1}(x)-\int_{0}^{x}K(x,y)p_{1}(y)dy,
\end{equation}and
\begin{equation}\label{rt}
r=q-\frac{\lambda}{2\varepsilon}.
\end{equation}\end{prop}

\textit{Proof:} The proof is very similar to that of Proposition 2 in \cite{rathnayake2020observer} and hence omitted. \hfill $\qed$

The inverse transformation of \eqref{ctbt} can be shown to be as follows:\begin{equation}\label{ink}
\hat{u}(x,t)=\hat{w}(x,t)+\int_{0}^{x}L(x,y)\hat{w}(y,t)dy,
\end{equation}where
\begin{equation}\label{Lsol}
L(x,y)=-\frac{\lambda}{\varepsilon}y\frac{J_{1}\big(\sqrt{\lambda(x^{2}-y^{2})/\varepsilon}\big)}{\sqrt{\lambda(x^{2}-y^{2})/\varepsilon}},
\end{equation}
for $0\leq y\leq x\leq 1$.

\begin{prop}\label{prop3}
Subject to Assumption \ref{ass1}, the closed-loop system which consists of the plant \eqref{ctp}  and the observer \eqref{cto} with the continuous-time control law \eqref{ctcl}, is globally exponentially stable in $L^{2}$-sense.
\end{prop}

\textit{Proof:} The proof is very similar to that of Proposition 3 in \cite{rathnayake2020observer} and hence omitted. \hfill $\qed$

\subsection{Emulation of the Observer-based Backstepping Boundary Control}
We aim to stabilize the closed-loop system containing the plant \eqref{ctp} and the observer \eqref{cto} while sampling the continuous-time controller $U(t)$ given by \eqref{ctcl} at a certain sequence of time instants $(t_{j})_{j\in\mathbb{N}}$. These sampling instants will be fully characterized later on for both sampled-data control (via  a maximum upper diameter of the sampling schedule) and event-triggered control (via an event trigger). The control input is held constant between two consecutive time instants. Therefore, we define the control input for $t\in[t_{j},t_{j+1}),j\in\mathbb{N}$ as
\begin{equation}\label{etcl}
U_{j}:=U(t_{j})=\int_{0}^{1}k(y)\hat{u}(y,t_{j})dy,
\end{equation}
where
\begin{equation}\label{frk1}
k(y):=rK(1,y)+K_{x}(1,y).
\end{equation}
Accordingly, the boundary conditions \eqref{ctpe3} and \eqref{ctoe3} are modified, respectively, as follows:\begin{equation}\label{mctpe3}
u_{x}(1,t)+qu(1,t)=U_{j},
\end{equation}\begin{equation}\label{mctoe3}
\hat{u}_{x}(1,t)+q\hat{u}(1,t)=U_{j}+\frac{\lambda}{2\varepsilon}\tilde{u}(1,t).
\end{equation}The deviation between the continuous-time control law and its sampled counterpart, referred to as the input holding error, is defined as follows:\begin{equation}\label{dt}
\begin{split}
d(t):=\int_{0}^{1}k(y)\big(\hat{u}(y,t_{j})-\hat{u}(y,t)\big)dy,
\end{split}
\end{equation}for $t\in[t_{j},t_{j+1}),j\in\mathbb{N}$. It can be shown that the backstepping transformation \eqref{ctbt}, applied on the system \eqref{ctoe1},\eqref{ctoe2},\eqref{mctoe3} between $t_{j}$ and $t_{j+1},j\in\mathbb{N}$, yields the following target system, valid for $t\in[t_{j},t_{j+1}),j\in\mathbb{N}$:
\begin{subequations}\label{ettsm}\begin{align}\label{ettsm1}
\hat{w}_{t}(x,t)&=\varepsilon \hat{w}_{xx}(x,t)+g(x)\tilde{w}(1,t),
\\\label{ettsm2}
\hat{w}(0,t)&=0,
\\\label{ettsm3}
\hat{w}_{x}(1,t)&=-r\hat{w}(1,t)+d(t)+\frac{\lambda}{2\varepsilon}\tilde{w}(1,t),
\end{align}\end{subequations}
where $g(x)$ and $r$ are given by \eqref{gt} and \eqref{rt}, respectively.

It is straightforward to see that the observer error system $\tilde{u}$ for $t\in[t_{j},t_{j+1}),j\in\mathbb{N}$ under the modified boundary conditions \eqref{mctpe3} and \eqref{mctoe3} will still be the same as \eqref{ctoe}. Therefore, the application of  the backstepping transformation \eqref{ctobt} on $\tilde{u}$ between $t_{j}$ and $t_{j+1},j\in\mathbb{N}$ yields the following observer error target system
\begin{subequations}\label{etoet}
\begin{align}\label{etoet1}
\tilde{w}_{t}(x,t)&=\varepsilon \tilde{w}_{xx}(x,t),
\\\label{etoet2}
\tilde{w}(0,t)&=0,
\\\label{etoet3}
\tilde{w}_{x}(1,t)&=-q\tilde{w}(1,t),
\end{align}\end{subequations}
valid for $t\geq 0$. 

\subsection{Well-posedness Issues}

\begin{prop}\label{cor1}
For given initial data $u[t_{j}],\hat{u}[t_{j}]\in L^{2}(0,1)$, there exist unique mappings $u,\hat{u}\in C^{0}([t_{j},t_{j+1}]; L^{2}(0,1))\cap C^{1}((t_{j},t_{j+1})\times [0,1])$ with $u[t],\hat{u}[t]\in C^{2}([0,1])$ which satisfy \eqref{ctpe2},\eqref{ctoe2},\eqref{etcl}-\eqref{mctoe3} for $t\in (t_{j},t_{j+1}]$ and \eqref{ctpe1}, \eqref{ctoe1} for $t\in (t_{j},t_{j+1}]$, $x\in(0,1)$.    
\end{prop}

\textit{Proof:} The proof is very similar to that of Proposition 4 in \cite{rathnayake2020observer} and hence omitted. \hfill $\qed$

\begin{cllry}\label{corf}
Let $\{t_j\geq 0,j=0,1,2,\ldots\}$ be an increasing sequence  of sampling times with $t_{0}=0,\lim_{j\rightarrow+\infty}(t_{j})=+\infty$. Then for every given initial data $u[0],\hat{u}[0]\in L^{2}(0,1)$, there exist unique mappings \mbox{$u,\hat{u}\in C^{0}(\mathbb{R_{+}};L^{2}(0,1))\cap C^{1}(I\times [0,1])$} with $u[t],\hat{u}[t]\in C^{2}([0,1])$ which satisfy \eqref{ctpe2},\eqref{ctoe2},\eqref{etcl}-\eqref{mctoe3} for all $t>0$ and \eqref{ctpe1}, \eqref{ctoe1} for all $t>0$, $x\in(0,1)$, where $I=\mathbb{R_{+}}\text{\textbackslash}\{t_{j}\geq 0,j\in\mathbb{N}\}$. 
\end{cllry}

\textit{Proof:} This is a straightforward consequence of Proposition \ref{cor1} and Theorem 4.10 in \cite{karafyllis2019input}. The solutions are constructed iteratively between consecutive triggering times. \hfill $\qed$

\begin{rmk}\label{ici}
Note that when $u[0],\hat{u}[0]\in H^{1}(0,1)$, it follows that $u[0],\hat{u}[0]\in L^{2}(0,1)$. Thus, the same existence and uniqueness results presented in Corollary \ref{corf} hold when $u[0],\hat{u}[0]\in H^{1}(0,1)$.  
\end{rmk}

\section{Observer-Based Sampled-Data Boundary Control }
\begin{figure}
\centering
\includegraphics[scale=0.75]{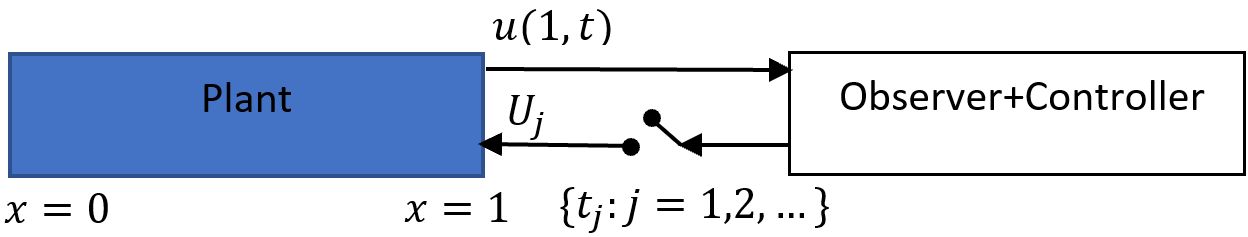}
\caption{Sampled-data observer-based closed-loop system.}
\end{figure}
In this section, we show the closed-loop exponential stability for the closed-loop system \eqref{ctpe1},\eqref{ctpe2},\eqref{ctoe1},\eqref{ctoe2},\eqref{etcl}-\eqref{mctoe3} under arbitrary sampling schedules of sufficiently small sampling period. The structure of the closed-loop system consisting of the plant and the observer-based controller is illustrated in Fig. 1. We present the main result in Theorem \ref{jl} and state several important estimates in Lemma \ref{slem1}-\ref{slem3}, which are required in proving Theorem \ref{jl}. Below are some definitions that we frequently use in this section. 

\begin{dfn}\label{SLdfn1}
For $\omega\in\mathbb{R},\theta>0$, consider the Sturm-Liouville (SL) operator $_{\theta}^{\omega}G:D_{\theta}\rightarrow L^{2}(0,1)$ defined as\begin{equation}\label{ASL1}
(_{\theta}^{\omega} Gf_{\theta})(x):=-\varepsilon f_{\theta}''(x)-\omega f_{\theta}(x),
\end{equation}for all $f_{\theta}\in D_{\theta}$ and $x\in (0,1)$ where $D_{\theta}\subseteq H^{2}(0,1)$ is given by\begin{equation}\label{ASL2}
D_{\theta}:=\{f_{\theta}\in H^{2}(0,1)\big |f_{\theta}(0)=f'_{\theta}(1)+\theta f_{\theta}(1)=0\}.
\end{equation}The eigenfunctions $\phi_{\theta,n}(x),n=1,2,\cdots$ of the SL operator $^\omega_\theta G$ defined by \eqref{ASL1} and \eqref{ASL2} are
\begin{equation}\label{egfn1}
\phi_{\theta,n}(x)=\Bigg(\frac{2\theta}{\theta+\cos^{2}(\nu_{\theta,n})}\Bigg)^{\frac{1}{2}}\sin(\nu_{\theta,n}x),
\end{equation}where $\nu_{\theta,n}>0,n=1,2,\cdots$ satisfies 
\begin{equation}\label{transcot}
\nu_{\theta,n}\cot(\nu_{\theta,n})=-\theta.
\end{equation}
The corresponding eigenvalues $_\theta^\omega\mu_{n},n=1,2,\cdots$ are given by\begin{equation}\label{egvl1}
_\theta^\omega\mu  _{n}=\varepsilon \nu_{\theta,n}^{2}-\omega.
\end{equation}The eigenfunctions $\phi_{\theta,n}(x),n=1,2,\cdots$ satisfy $\phi_{\theta,n}(x)\in D_{\theta},(_{\theta}^{\omega}G\phi_{\theta,n})(x)={_\theta^\omega}\mu_{n}\phi_{\theta,n}(x)$ 
and form  an orthonomal basis of $L^{2}(0,1)$. Furthermore, the eigenvalues form an infinite, increasing sequence  ${_\theta^\omega}\mu_{1}<{_\theta^\omega}\mu_{2}\cdots<{_\theta^\omega}\mu_{n}\cdots$ with $\lim_{n\rightarrow+\infty}({_\theta^\omega}\mu_{n})=+\infty$.
\end{dfn}

\begin{dfn}\label{SLdfn2}
Consider the Sturm-Liouville (SL) operator $\tilde{G}:\tilde{D}\rightarrow L^{2}(0,1)$ is defined as\begin{equation}\label{GSL1}
(\tilde{G}\tilde{f})(x):=-\varepsilon \tilde{f}''(x)+2\varepsilon q\tilde{f}(x),
\end{equation}for all $\tilde{f}\in \tilde{D}$ and $x\in (0,1)$ where $\tilde{D}\subseteq H^{2}(0,1)$ is given by\begin{equation}\label{GSL2}
\tilde{D}:=\{\tilde{f}\in H^{2}(0,1)\big |\tilde{f}'(0)=\tilde{f}(1)=0\}.
\end{equation}The eigenfunctions $\tilde{\phi}_{n}(x),n=1,2,\cdots$ of the SL operator $\tilde{G}$ defined by \eqref{GSL1} and \eqref{GSL2} are \begin{equation}
\tilde{\phi}_{n}(x)=\sqrt{2}\cos\big((n-\frac{1}{2})\pi x\big),
\end{equation}with the corresponding eigenvalues $\tilde{\mu}_{n}>0,n=1,2,\cdots$ given by
\begin{equation}
\tilde{\mu}_{n}=\varepsilon\big(n-\frac{1}{2}\big)^{2}\pi^{2}+2\varepsilon q,
\end{equation}
such that $\tilde{\phi}_{n}(x)\in\tilde{D},(\tilde{G}\tilde{\phi}_{n})(x)=\tilde{\mu}_n\tilde{\phi}_{n}(x)$, and $\tilde{\phi}_{n}(x),n=1,2,\cdots$ form an orthonomal basis of $L^{2}(0,1)$. Furthermore, the eigenvalues form an infinite, increasing sequence $0<\tilde{\mu}_{1}<\tilde{\mu}_{2}\cdots<\tilde{\mu}_{n}\cdots$ with $\lim_{n\rightarrow+\infty}(\tilde{\mu}_{n})=+\infty$.
\end{dfn}

\begin{lemm}\label{slem1}
For every $\tilde{w}[0]\in H^{1}(0,1)$, the unique solution $\tilde{w}\in C^{0}(\mathbb{R_{+}};L^{2}(0,1))$  obeying \eqref{etoet2},\eqref{etoet3} for all $t>0$ and \eqref{etoet1} for all $t>0$, $x\in(0,1)$, satisfy the following estimates for all $q>0$ and $t\geq 0$\begin{equation}\label{esc1}
\Vert\tilde{w}_{x}[t]\Vert\leq \big(\Vert\tilde{w}_{x}[0]\Vert+M_{1}\Vert\tilde{w}[0]\Vert\big)e^{-\sigma_{1} t},
\end{equation} 
\begin{equation}\label{esc2}
\Vert\tilde{w}[t]\Vert\leq \Vert\tilde{w}[0]\Vert e^{-\sigma_{1} t},
\end{equation}where\begin{equation}\label{phmk1}
M_{1}=2q+\frac{2\varepsilon q^{2}}{\tilde{\mu}_{1}-\sigma_{1}}\text{ and }\sigma_{1}\in\big[0,\min(\tilde{\mu}_{1},{_q^0}\mu_{1})\big).
\end{equation} Here  $\tilde{\mu}_{1}=\frac{\varepsilon\pi^{2}}{4}+2\varepsilon q$ is the smallest eigenvalue of the SL operator $\tilde{G}:\tilde{D}\rightarrow L^{2}(0,1)$ defined in Definition \ref{SLdfn2}. ${_q^0}\mu_{1}=\varepsilon \nu^{2}_{q,1}$, where  $\nu_{q,1}\in (\pi/2,\pi)$ satisfies \eqref{transcot} with $\theta=q$, is the smallest eigenvalue of a SL operator ${_q^0}G:D_{q}\rightarrow L^{2}(0,1)$ befitting Definition \ref{SLdfn1}. 
\end{lemm}

\textit{Proof:} By the straightforward application of Theorem 5.8 in \cite{karafyllis2019input}, we can obtain the estimates \eqref{esc1}-\eqref{phmk1}. \hfill $\qed$

\begin{lemm}\label{slem2}
For every increasing sequence $\{t_{j}\geq 0,j=1,2,\ldots\}$ with $t_{0}=0,\lim_{j\rightarrow+\infty}(t_{j})=+\infty$ and for every $\hat{w}[0]\in L^{2}(0,1)$, the unique mapping $\hat{w}\in C^{0}(\mathbb{R_{+}};L^{2}(0,1))\cap C^{1}(I\times [0,1])$ with $\hat{w}[t]\in C^{2}([0,1])$ obeying \eqref{ettsm2},\eqref{ettsm3} for all $t>0$ and \eqref{ettsm1} for all $t>0$, $x\in(0,1)$, where $I=\mathbb{R_{+}}\text{\textbackslash}\{t_{j}\geq 0,j\in\mathbb{N}\}$, satisfies the following estimate for all $r>0$ and all $t\geq 0$\begin{equation}\label{esc3}
\begin{split}
\Vert\hat{w}[t]\Vert\leq &\Vert\hat{w}[0]\Vert e^{-\sigma_{2}t}+C_{1}\sup_{0\leq s\leq t}\Big(\vert d(s)\vert e^{-\sigma_{2}(t-s)}\Big)\\&+C_{2}\sup_{0\leq s\leq t}\Big(\vert\tilde{w}(1,s)\vert e^{-\sigma_{2}(t-s)}\Big),
\end{split}
\end{equation}where\begin{equation}\label{mmd1}
\sigma_{2}\in[0,{_r^0}\mu_{1}),
\end{equation}
\begin{equation}\label{mmd2}
C_{1}=\frac{{_r^0}\mu_{1}}{\sqrt{3}(1+r)({_r^0}\mu_{1}-\sigma_{2})},
\end{equation}
\begin{equation}\label{mmd3}
C_{2}=\frac{{_r^0}\mu_{1}\lambda+2\sqrt{3}\varepsilon(1+r)\Vert g\Vert}{2\sqrt{3}\varepsilon(1+r)({_r^0}\mu_{1}-\sigma_{2})}.
\end{equation}Here  ${_r^0}\mu_{1}=\varepsilon \nu^{2}_{r,1}$, where  $\nu_{r,1}\in (\pi/2,\pi)$ satisfies \eqref{transcot} with $\theta=r$, is the smallest eigenvalue of a SL operator ${_r^0}G:D_{r}\rightarrow L^{2}(0,1)$ befitting Definition \ref{SLdfn1}. 
\end{lemm}

\textit{Proof:} For every increasing sequence $\{t_{j}\geq 0,j=1,2,\ldots\}$ with $t_{0}=0,\lim_{j\rightarrow+\infty}(t_{j})=+\infty$ and for every $\hat{w}[0]\in L^{2}(0,1)$, the existence/uniqueness of the mapping $\hat{w}\in C^{0}(\mathbb{R_{+}};L^{2}(0,1))\cap C^{1}(I\times [0,1])$ with $\hat{w}[t]\in C^{2}([0,1])$ satisfying \eqref{ettsm2},\eqref{ettsm3} for all $t>0$ and \eqref{ettsm1} for all $t>0$, $x\in(0,1)$, where $I=\mathbb{R_{+}}\text{\textbackslash}\{t_{j}\geq 0,j\in\mathbb{N}\}$ is guaranteed by Corollary \ref{corf} due to the transformation \eqref{ctbt}. Furthermore, we have that $\int_{0}^{1}\tilde{w}_{x}(x,t)dx=\tilde{w}(1,t)$ as $\tilde{w}(0,t)=0$. Hence, it follows from Cauchy-Schwarz inequality that $\vert\tilde{w}(1,t)\vert\leq \Vert\tilde{w}_{x}[t]\Vert$ for all $t\geq 0$. But Lemma \ref{slem1} establishes the boundedness of $\Vert\tilde{w}_{x}[t]\Vert$ which in turn ensures the boundedness of $\vert\tilde{w}(1,t)\vert$. Therefore, by the straightforward application of Theorem 5.3 in \cite{karafyllis2019input}, we can obtain the estimate \eqref{esc3}-\eqref{mmd3}. \hfill $\qed$ 

\begin{lemm}\label{slem3}
Consider the SL operator ${_q^\lambda}G: D_q\rightarrow L^{2}(0,1)$ befitting Definition \ref{SLdfn1} whose eigenfunctions $\phi_{q,n}$ and the corresponding eigenvalues ${_q^\lambda}\mu_n$ respectively satisfy \eqref{egfn1} and \eqref{egvl1} with $\theta=q$ and $\omega=\lambda$ for $n=1,2,\ldots$. Then, for any $T>0$ such that $\sup_{j\geq 0}(t_{j+1}-t_{j})\leq T$ with $t_0=0$ and $\lim_{j\rightarrow +\infty}(t_j)=+\infty,$ the following estimate holds for all $t\geq 0$
\begin{equation}\label{esd}
\begin{split}
&\vert d(t)\vert e^{\sigma t}\\&\leq \tilde{L}Te^{\sigma T}\sum_{n=1}^{N}\Big(\varepsilon\Vert k\Vert\big\vert k_{n}\phi_{q,n}(1)\big\vert+\vert {_q^\lambda}\mu_{n}k_{n}\vert\Big)\sup_{0\leq s\leq t}\Big(\Vert\hat{w}[s]\Vert e^{\sigma s}\Big)\\&
+\tilde{L}\Vert k-h\Vert\Big(e^{\sigma T}+1\Big)\sup_{0\leq s\leq t}\Big(\Vert\hat{w}[s]\Vert e^{\sigma s}\Big)
\\&
+\frac{Te^{\sigma T}}{\sqrt{2}}\sum_{n=1}^{N}\Big(\frac{\lambda}{2}\vert k_{n}\phi_{q,n}(1)\vert+\Vert p_{1}\Vert\vert k_{n}\vert\Big)\sup_{0\leq s\leq t}(\Vert\tilde{w}[s])\Vert e^{\sigma s})\\&
+\frac{Te^{\sigma T}}{\sqrt{2}}\sum_{n=1}^{N}\Big(\frac{\lambda}{2}\vert k_{n}\phi_{q,n}(1)\vert+\Vert p_{1}\Vert\vert k_{n}\vert\Big)\sup_{0\leq s\leq t}(\Vert\tilde{w}_{x}[s])\Vert e^{\sigma s}),
\end{split}
\end{equation}where $d(t)$ is the input holding error given by \eqref{dt}, $N\geq 1,$ $\sigma>0$
\begin{equation}\label{L1}
\tilde{L}:=1+\Big(\int_{0}^{1}\int_{0}^{x}L^{2}(x,y)dydx\Big)^{1/2},
\end{equation}\begin{equation}\label{eh}
h(x):=\sum_{n=1}^{N}k_{n}\phi_{q,n}(x),
\end{equation}for all $x\in[0,1]$, and
\begin{equation}
k_{n}:=\int_{0}^{1}k(y)\phi_{q,n}(y)dy.
\end{equation}
The constants $\sigma_1$ and $\sigma_{2}$ are defined in \eqref{phmk1} and \eqref{mmd1}, respectively whereas $p_1(x),L(x,y),\text{ and }k(y)$ are given by \eqref{p1},\eqref{Lsol}, and \eqref{frk1}, respectively.
\end{lemm}

\textit{Proof:}
Let us define\begin{equation}\label{dbart}
\bar{d}(t):=\int_{0}^{1}h(y)\big(\hat{u}(y,t_{j})-\hat{u}(y,t)\big)dy,
\end{equation}for $t\in[t_{j},t_{j+1}),j\in\mathbb{N}$. The definitions \eqref{dt} and \eqref{dbart} imply that\begin{equation}\label{dtnw}
d(t)=\bar{d}(t)+\int_{0}^{1}\big(k(y)-h(y)\big)\big(\hat{u}(y,t_{j})-\hat{u}(y,t)\big)dy,
\end{equation}
for $t\in[t_{j},t_{j+1}),j\in\mathbb{N}$. Differentiating \eqref{dbart} w.r.t time in $t\in(t_{j},t_{j+1}),j\in\mathbb{N},$ using \eqref{ctoe1}, and integrating by parts twice, we can obtain that\begin{equation}
\begin{split}
\dot{\bar{d}}(t)&=-\int_{0}^{1}h(y)\hat{u}_{t}(y,t)dy\\
&=-\lambda\int_{0}^{1}h(y)\hat{u}(y,t)dy-\int_{0}^{1}h(y)p_{1}(y)dy\tilde{u}(1,t)\\&\quad
-\varepsilon h(1)\hat{u}_{x}(1,t)+\varepsilon h(0)\hat{u}_{x}(0,t)+\varepsilon h'(1)\hat{u}(1,t)\\&\quad-
\varepsilon h'(0)\hat{u}(0,t)-\varepsilon\int_{0}^{1}h''(y)\hat{u}(y,t)dy,
\end{split}
\end{equation}
As $\phi_{q,n}\in D_q$ for all $n=1,2,\ldots,$ it follows from \eqref{eh} that $h(x)\in D_q$. Thus, from Definition \ref{SLdfn1}, we have that $({_q^\lambda }Gh)(x)=-\varepsilon h''(x)-\lambda h(x)$ and $h(0)=h'(1)+q h(1)=0.$ Therefore, recalling \eqref{ctoe2} and \eqref{mctoe3}, we can obtain that
\begin{equation}\label{ddbart}
\begin{split}
\dot{\bar{d}}(t)=&-\varepsilon h(1)U(t_{j})+\int_{0}^{1}(_{q}^{\lambda}Gh)(y)\hat{u}(y,t)dy\\&-\frac{\lambda}{2}h(1)\tilde{u}(1,t)
-\int_{0}^{1}h(y)p_{1}(y)dy\tilde{u}(1,t),
\end{split}
\end{equation}for   $t\in(t_{j},t_{j+1}),j\in\mathbb{N}$. Using \eqref{etcl}, recalling that $(_{q}^{\lambda}G\phi_{q,n})(x)={_q^\lambda}\mu_{n}\phi_{q,n}(x)$ from Definition \ref{SLdfn1}, and using \eqref{eh}, we can rewrite \eqref{ddbart} as
\begin{equation}
\begin{split}
&\dot{\bar{d}}(t)=-\varepsilon\sum_{n=1}^{N}k_{n}\phi_{q,n}(1)\int_{0}^{1}k(y)\hat{u}(y,t_{j})dy\\&
+\sum_{n=1}^{N}{_q^\lambda}\mu_{n}k_{n}\int_{0}^{1}\phi_{q,n}(y)\hat{u}(y,t)dy
\\&
-\sum_{n=1}^{N}\Big(\frac{\lambda}{2}k_{n}\phi_{q,n}(1)+k_{n}\int_{0}^{1}\phi_{q,n}(y)p_{1}(y)dy\Big)\tilde{u}(1,t),
\end{split}
\end{equation}
for  $t\in(t_{j},t_{j+1}),j\in\mathbb{N}$. Further, using Cauchy-Schwarz inequality and noting $\Vert\phi_{q,n}\Vert=1$ from Definition \ref{SLdfn1}, we can show that
\begin{equation}\label{alc}
\begin{split}
\vert\dot{\bar{d}}(t)\vert\leq &\varepsilon\Vert k\Vert\Vert\hat{u}[t_{j}]\Vert\sum_{n=1}^{N}\big\vert k_{n}\phi_{q,n}(1)\big\vert+\Vert\hat{u}[t]\Vert\sum_{n=1}^{N}\vert{_q^\lambda}\mu_{n} k_{n}\vert\\&
+\vert\tilde{u}(1,t)\vert\sum_{n=1}^{N}\Big(\frac{\lambda}{2}\vert k_{n}\phi_{q,n}(1)\vert+\Vert p_{1}\Vert\vert k_{n}\vert\Big),
\end{split}
\end{equation}
for  $t\in(t_{j},t_{j+1}),j\in\mathbb{N}$. Therefore, for a $T>0$ such that $\sup_{j\geq 0}(t_{j+1}-t_{j})\leq T$, let us integrate \eqref{alc} in $t\in[t_{j},t_{j+1}),j\in\mathbb{N}$. Then, noting $\bar{d}(t_j)=0$ from \eqref{dbart},  we get
\begin{equation}\label{aup1}
\begin{split}
&\vert\bar{d}(t)\vert\\&\leq T\varepsilon\Vert k\Vert\Vert\hat{u}[t_{j}]\Vert\sum_{n=1}^{N}\big\vert k_{n}\phi_{q,n}(1)\big\vert+T\sup_{t_{j}\leq s\leq t}(\Vert\hat{u}[s]\Vert)\sum_{n=1}^{N}\vert {_q^\lambda}\mu_{n}k_{n}\vert\\&
+T\sup_{t_{j}\leq s\leq t}(\vert\tilde{u}(1,s)\vert)\sum_{n=1}^{N}\Big(\frac{\lambda}{2}\vert k_{n}\phi_{q,n}(1)\vert+\Vert p_{1}\Vert\vert k_{n}\vert\Big),
\end{split}
\end{equation}
for $t\in[t_{j},t_{j+1}),j\in\mathbb{N}$. Furthermore, using \eqref{aup1} along with the fact that $\sup_{j\geq 0}(t_{j+1}-t_{j})\leq T$, we can show
\begin{equation}
\begin{split}
&\vert\bar{d}(t)\vert e^{\sigma t}\\&\leq Te^{\sigma T}\sum_{n=1}^{N}\Big(\varepsilon\Vert k\Vert\big\vert k_{n}\phi_{q,n}(1)\big\vert+\vert k_{n}{_q^\lambda}\mu_{n}\vert\Big)\sup_{0\leq s\leq t}\Big(\Vert\hat{u}[s]\Vert e^{\sigma s}\Big)\\&
+Te^{\sigma T}\sum_{n=1}^{N}\Big(\frac{\lambda}{2}\vert k_{n}\phi_{q,n}(1)\vert+\Vert p_{1}\Vert\vert k_{n}\vert\Big)\sup_{0\leq s\leq t}(\vert\tilde{u}(1,s)\vert e^{\sigma s}),
\end{split}
\end{equation}
for any $\sigma>0$. Thus, we can show recalling \eqref{dtnw} that
\begin{equation}\label{aup2}
\begin{split}
&\vert d(t)\vert e^{\sigma t}\\&\leq Te^{\sigma T}\sum_{n=1}^{N}\Big(\varepsilon\Vert k\Vert\big\vert k_{n}\phi_{q,n}(1)\big\vert+\vert {_q^\lambda}\mu_{n}k_{n}\vert\Big)\sup_{0\leq s\leq t}\Big(\Vert\hat{u}[s]\Vert e^{\sigma s}\Big)\\&
+Te^{\sigma T}\sum_{n=1}^{N}\Big(\frac{\lambda}{2}\vert k_{n}\phi_{q,n}(1)\vert+\Vert p_{1}\Vert\vert k_{n}\vert\Big)\sup_{0\leq s\leq t}(\vert\tilde{u}(1,s)\vert e^{\sigma s})\\&
+\Vert k-h\Vert\Vert\hat{u}[t_{j}]\Vert e^{\sigma t}+\Vert k-h\Vert\Vert\hat{u}[t]\Vert e^{\sigma t},
\end{split}
\end{equation}
for $t\in [t_{j},t_{j+1}),j\in\mathbb{N}$. Using Cauchy-Schwarz inequality and \eqref{ink}, we can obtain that $\Vert\hat{u}[t]\Vert\leq\tilde{L}\Vert\hat{w}[t]\Vert$ where $\tilde{L}$ is given by \eqref{L1}. Furthermore, it follows from \eqref{ctobt} that $\tilde{w}(1,t)=\tilde{u}(1,t)$. Therefore, using \eqref{aup2} and noting the fact that $\sup_{j\geq 0}(t_{j+1}-t_{j})\leq T$, we get 
\begin{equation}\label{ne1}
\begin{split}
&\vert d(t)\vert e^{\sigma t}\\&\leq \tilde{L}Te^{\sigma T}\sum_{n=1}^{N}\Big(\varepsilon\Vert k\Vert\big\vert k_{n}\phi_{q,n}(1)\big\vert+\vert {_q^\lambda}\mu_{n}k_{n}\vert\Big)\sup_{0\leq s\leq t}\Big(\Vert\hat{w}[s]\Vert e^{\sigma s}\Big)\\&
+\tilde{L}\Vert k-h\Vert\Big(e^{\sigma T}+1\Big)\sup_{0\leq s\leq t}\Big(\Vert\hat{w}[s]\Vert e^{\sigma s}\Big)
\\&
+Te^{\sigma T}\sum_{n=1}^{N}\Big(\frac{\lambda}{2}\vert k_{n}\phi_{q,n}(1)\vert+\Vert p_{1}\Vert\vert k_{n}\vert\Big)\sup_{0\leq s\leq t}(\vert\tilde{w}(1,s)\vert e^{\sigma s}),
\end{split}
\end{equation}for all $t\geq 0$. But, using Agmon's and Young's inequalities along with the fact that $\tilde{w}(0,t)=0$, we can obtain that
\begin{equation}\label{ne2}
\vert \tilde{w}(1,t)\vert\leq \frac{1}{\sqrt{2}}\Vert\tilde{w}[t]\Vert+\frac{1}{\sqrt{2}}\Vert\tilde{w}_{x}[t]\Vert,
\end{equation}
for all $t\geq 0$. Thus, the estimate \eqref{esd} directly follows from \eqref{ne1} and \eqref{ne2}. \hfill $\qed$  

With Lemma \ref{slem1}-\ref{slem3} at hand, we are now in position to prove the following result, which shows that the observer-based boundary sampled-data feedback control law obtained by the emulation
of the continuous-time observer-based backstepping design works.
\begin{thme}\label{jl}
There exist constants $\sigma,T^{*}>0$, $M:=M(T^*)>0$, and increasing sequences $\{t_{j}\geq 0,j=1,2,\ldots\}$ with $t_{0}=0$, $\sup_{j\geq 0}(t_{j+1}-t_{j})\leq T^{*},\text{ and }\lim_{j\rightarrow+\infty}(t_{j})=+\infty$ such that, for every $u[0],\hat{u}[0]\in H^{1}(0,1)$ and subject to Assumption \ref{ass1} and the controller \eqref{etcl},\eqref{frk1}, the closed-loop system consisting of the unique mappings $u,\hat{u}\in C^{0}(\mathbb{R_{+}};L^{2}(0,1))\cap C^{1}(I\times [0,1])$ with $u[t],\hat{u}[t]\in C^{2}([0,1])$ obeying \eqref{ctpe2},\eqref{ctoe2},\eqref{mctpe3},\eqref{mctoe3} for all $t>0$ and \eqref{ctpe1}, \eqref{ctoe1} for all $t>0$, $x\in(0,1)$, where $I=\mathbb{R_{+}}\text{\textbackslash}\{t_{j}\geq 0,j\in\mathbb{N}\}$, satisfy the following estimate\begin{equation}
\begin{split}
\Vert\hat{u}[t]\Vert&+\Vert\tilde{u}[t]\Vert+\Vert\tilde{u}_{x}[t]\Vert\\&\leq M(T^{*})\big(\Vert\hat{u}[0]\Vert+\Vert\tilde{u}[0]\Vert+\Vert\tilde{u}_{x}[0]\Vert\big)e^{-\sigma t},
\end{split}
\end{equation}
for all $t\geq 0$. Here $\tilde{u}[t]=u[t]-\hat{u}[t]$. In other words, the closed-loop system is globally exponentially stable in the sense of the norm $\Vert\hat{u}[t]\Vert+\Vert\tilde{u}[t]\Vert+\Vert\tilde{u}_{x}[t]\Vert.$
\end{thme}

\textit{Proof:} Let $T^{*}>0$ be a constant to be selected. Furthermore, let an increasing sequence $\{t_{j}\geq 0,j=1,2,\ldots\}$ with $t_{0}=0,\text{ }\sup_{j\geq 0}(t_{j+1}-t_{j})\leq T^{*},\lim_{j\rightarrow+\infty}(t_{j})=+\infty$ and $u[0],\hat{u}[0]\in H^{1}(0,1)$ be given. Then the existence/uniqueness of the mappings $u,\hat{u}\in C^{0}(\mathbb{R_{+}};L^{2}(0,1))\cap C^{1}(I\times [0,1])$ with $u[t],\hat{u}[t]\in C^{2}([0,1])$ satisfying \eqref{ctpe2},\eqref{ctoe2},\eqref{etcl}-\eqref{mctoe3} for all $t>0$ and \eqref{ctpe1}, \eqref{ctoe1} for all $t>0$, $x\in(0,1)$ is guaranteed by Corollary \ref{corf} and Remark \ref{ici}. 

Before proceeding with the proof of the exponential stability of the closed-loop system, we present several definitions. Let us define \begin{equation}\label{jl1}
\tilde{P}:=1+\Big(\int_{0}^{1}\int_{x}^{1}P^{2}(x,y)dydx\Big)^{1/2},
\end{equation}
\begin{equation}\label{jl2}
\tilde{Q}:=1+\Big(\int_{0}^{1}\int_{x}^{1}Q^{2}(x,y)dydx\Big)^{1/2},
\end{equation}
\begin{equation}\label{jl3}
\tilde{K}:=1+\Big(\int_{0}^{1}\int_{0}^{x}K^{2}(x,y)dydx\Big)^{1/2},
\end{equation}where $P(x,y),Q(x,y),\text{ and }K(x,y)$are given by \eqref{solP},\eqref{solQ}, and \eqref{ctcks}, respectively, and define
\begin{equation}\label{jnna}
\sigma:=\min\{\sigma_1,\sigma_2\},
\end{equation}
where $\sigma_1$ and $\sigma_2$ are defined in \eqref{phmk1} and \eqref{mmd1}, respectively. Let us select $N\geq 1$ in \eqref{eh} sufficiently large so that 
\begin{equation}\label{cndm}
2C_1\tilde{L}\Vert k-h\Vert<1, 
\end{equation}where $C_{1}$ and $\tilde{L}$ are given by \eqref{mmd2} and \eqref{L1}, respectively. 

Noting that $\sigma_{2}\geq\sigma$ from \eqref{jnna} and using \eqref{ne2}, we can obtain from Lemma \ref{slem2} that
\begin{equation}\label{iss1}
\begin{split}
&\Vert\hat{w}[t]\Vert e^{\sigma t}\leq\Vert\hat{w}[0]\Vert +C_{1}\sup_{0\leq s\leq t}\Big(\vert d(s)\vert e^{\sigma s}\Big)\\&
+\frac{C_{2}}{\sqrt{2}}\sup_{0\leq s\leq t}\Big(\Vert\tilde{w}[s]\Vert e^{\sigma s}\Big)+\frac{C_{2}}{\sqrt{2}}\sup_{0\leq s\leq t}\Big(\Vert\tilde{w}_{x}[s]\Vert e^{\sigma s}\Big),
\end{split}
\end{equation}
for all $t\geq 0$. Note that above we have used the fact   $\sup_{0\leq s\leq t}\big(\vert d(s)\vert e^{-\sigma_2(t-s)}\big)\leq \sup_{0\leq s\leq t}\big(\vert d(s)\vert e^{-\sigma(t-s)}\big)$. Similarly, from Lemma \ref{slem1}, we can obtain 
\begin{equation}\label{aln1}
\Vert\tilde{w}_{x}[t]\Vert e^{\sigma t}\leq \Vert\tilde{w}_{x}[0]\Vert+M\Vert\tilde{w}[0]\Vert,
\end{equation}
\begin{equation}\label{aln2}
\Vert\tilde{w}[t]\Vert e^{\sigma t}\leq \Vert\tilde{w}[0]\Vert,
\end{equation}
for all $t\geq 0$ as $\sigma_1\geq \sigma$. Therefore, using \eqref{aln1} and \eqref{aln2}, we we get from \eqref{iss1} that 
\begin{equation}\label{aln3}
\begin{split}
\Vert\hat{w}[t]\Vert e^{\sigma t}\leq &\Vert\hat{w}[0]\Vert+\frac{C_{2}}{\sqrt{2}}(M+1)\Vert\tilde{w}[0]\vert
\\&+\frac{C_{2}}{\sqrt{2}}\Vert\tilde{w}_{x}[0]\Vert+C_{1}\sup_{0\leq s\leq t}\Big(\vert d(s)\vert e^{\sigma s}\Big)
\end{split},
\end{equation}
for all $t\geq 0$. By summing \eqref{aln1},\eqref{aln2},\eqref{aln3} together, we can derive the following estimate: 
\begin{equation}\label{aln4}
\begin{split}
&\sup_{0\leq s\leq t}(\Vert\hat{w}[s]\Vert e^{\sigma s})+\sup_{0\leq s\leq t}(\Vert\tilde{w}[s]\Vert e^{\sigma s})+\sup_{0\leq s\leq t}(\Vert \tilde{w}_{x}[s]\Vert e^{\sigma s})\\&\leq \Vert\hat{w}[0]\Vert+\Big(\frac{C_{2}}{\sqrt{2}}+1\Big)(M+1)\Vert\tilde{w}[0]\Vert+\Big(\frac{C_{2}}{\sqrt{2}}+1\Big)\Vert\tilde{w}_{x}[0]\Vert\\&+C_{1}\sup_{0\leq s\leq t}\Big(\vert d(s)\vert e^{\sigma s}\Big).
\end{split}
\end{equation}Therefore, using \eqref{esd} and \eqref{aln4}, we can show have that
\begin{equation}\label{aln5}
\begin{split}
&\gamma_{1}(T)\sup_{0\leq s\leq t}(\Vert\hat{w}[s]\Vert e^{\sigma s})+\gamma_{2}(T)\sup_{0\leq s\leq t}(\Vert\tilde{w}[s]\Vert e^{\sigma s})\\&+\gamma_{2}(T)\sup_{0\leq s\leq t}(\Vert \tilde{w}_{x}[s]\Vert e^{\sigma s})\\&\leq \Vert\hat{w}[0]\Vert+\Big(\frac{C_{2}}{\sqrt{2}}+1\Big)(M+1)\Vert\tilde{w}[0]\Vert+\Big(\frac{C_{2}}{\sqrt{2}}+1\Big)\Vert\tilde{w}_{x}[0]\Vert,
\end{split}
\end{equation}where
\begin{equation}\label{hn1}
\begin{split}
\gamma_{1}(T)=&1-C_{1}\tilde{L}Te^{\sigma T}\sum_{n=1}^{N}\Big(\varepsilon\Vert k\Vert\big\vert k_{n}\phi_{q,n}(1)\big\vert+\vert {_q^\lambda}\mu_{n}k_{n}\vert\Big)\\&-C_{1}\tilde{L}\Vert k-h\Vert\Big(e^{\sigma T}+1\Big),
\end{split}
\end{equation}
\begin{equation}\label{hn2}
\gamma_{2}(T)=1-\frac{C_{1}T}{\sqrt{2}}e^{\sigma T}\sum_{n=1}^{N}\Big(\frac{\lambda}{2}\vert k_{n}\phi_{q,n}(1)\vert+\Vert p_{1}\Vert\vert k_{n}\vert\Big).
\end{equation}
A Sampling Diameter $T^{*}$ should be chosen from the set
\begin{equation}\label{set1}
T^{*}=\{T^{*}>0|\gamma_{1}(T^{*})>0\wedge \gamma_{2}(T^{*})>0\}.
\end{equation}The existence of such $T^{*}$ is guaranteed. This can be shown as follows. We have that  $\gamma_{1}(0)=1-2C_{1}\tilde{L}\Vert k-h\Vert>0$ as $N\geq 1$ is chosen to satisfy \eqref{cndm}. We also have that $\gamma_{2}(0)=1$. Furthermore, it is easy to observe that $\gamma_{1}(T)$ and $\gamma_{2}(T)$ are both continuous in $T$. Therefore, \eqref{set1} cannot be a null set. 

Let $\Xi(T^{*})$ be defined as
\begin{equation}
\Xi(T^{*}):=\min\{\gamma_{1}(T^{*}),\gamma_{2}(T^{*})\}>0.
\end{equation}
Then,  we can obtain from \eqref{aln5} that 
\begin{equation}
\begin{split}
&\Vert\hat{w}[t]\Vert +\Vert\tilde{w}[t]\Vert +\Vert \tilde{w}_{x}[t]\Vert\\&\leq (\Xi(T^{*}))^{-1}\Big(\Vert\hat{w}[0]\Vert+(C_2/\sqrt{2}+1)(M+1)\Vert\tilde{w}[0]\Vert\\&\quad\quad\quad\quad\quad\quad+(C_2/\sqrt{2}+1)\Vert\tilde{w}_{x}[0]\Vert\Big)e^{-\sigma t}.
\end{split}
\end{equation}With the aid of Cauchy-Schwarz inequality and the transformations \eqref{ctobt},\eqref{ibtoe},\eqref{ctbt}, and \eqref{ink}, we can show that 
\begin{equation}
\begin{split}
\Vert\hat{u}[t]\Vert &+\Vert\tilde{u}[t]\Vert +\Vert \tilde{u}_{x}[t]\Vert\\&\leq \frac{\Omega_1\Omega_2}{\Xi(T^{*})}\Big(\Vert\hat{u}[0]\Vert+\Vert\tilde{u}[0]\Vert+\Vert\tilde{u}_{x}[0]\Vert\Big)e^{-\sigma t},
\end{split}
\end{equation}
where
\begin{equation}
\Omega_1=\max\bigg\{\tilde{L},\tilde{P}+\frac{\lambda}{2\varepsilon}+\sqrt{\int_{0}^{1}\int_{x}^{1}P_{x}^2(x,y)dydx},1\bigg\}>0,
\end{equation}
\begin{equation}
\begin{split}
\Omega_2=\max\Big\{&\tilde{K},(C_2/\sqrt{2}+1)\bigg((M+1)\tilde{Q}+\frac{\lambda}{2\varepsilon}\\&+\sqrt{\int_{0}^{1}\int_{x}^{1}Q_{x}^2(x,y)dydx}\bigg),C_2/\sqrt{2}+1\Big\}>0,
\end{split}
\end{equation}
with $\tilde{L},\tilde{P},\tilde{Q},$ and $\tilde{K}$ defined in \eqref{L1},\eqref{jl1}-\eqref{jl3}, respectively.\hfill $\qed$   
\begin{rmk}
Following from \eqref{set1}, the maximum upper diameter $T^{*}>0$ of the sampling schedule should satisfy the inequalities $\gamma_1(T^{*})>0$ and $\gamma_{2}(T^{*})>0$ where $\gamma_1(\cdot)$ and $\gamma_{2}(\cdot)$ are given by \eqref{hn1} and \eqref{hn2}, respectively. The solution space for the maximum upper diameter is conservative because it is obtained using small-gain arguments presented in the proof of Theorem \ref{jl}. However, it can be used to understand the qualitative dependence of the maximum upper
diameter of the sampling schedule on the control Kernel $K$ and the system parameters.  
\end{rmk}
\section{Observer-Based Event-Triggered Boundary Control}

\begin{figure}
\centering
\includegraphics[scale=0.75]{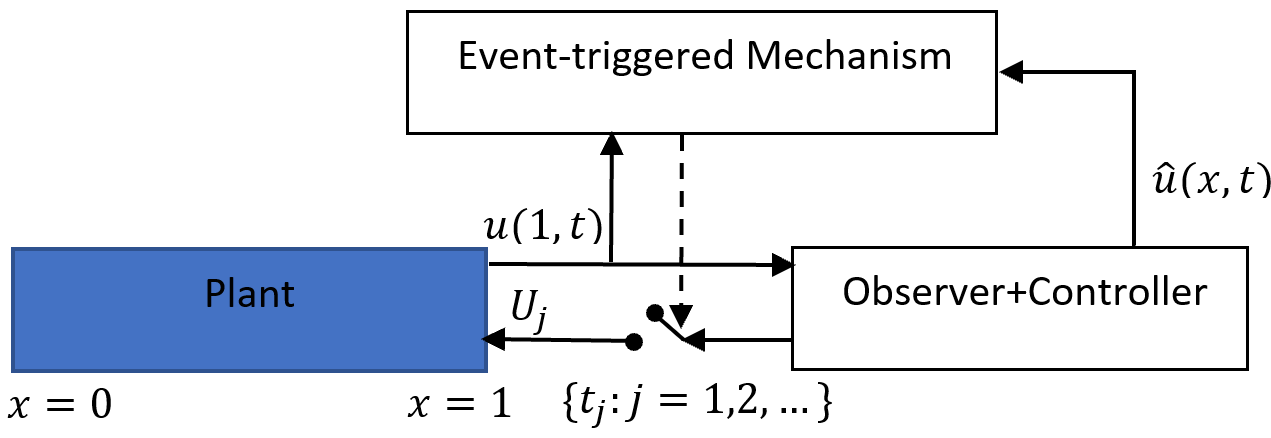}
\caption{Event-triggered observer-based closed-loop system.}
\end{figure}

Now we present the observer-based event-triggered boundary control approach considered in this work. The closed-loop
system consisting of the plant, the observer-based controller, and the event trigger is depicted in Fig. 2. The event-triggering condition involves the square of the input holding error $d(t)$ and a dynamic variable $m(t)$ that depends on the information of the systems \eqref{ettsm} and \eqref{etoet}.

\begin{dfn}\label{def1}
Let $\eta,\gamma,\rho,\beta_{1},\beta_{2},\beta_{3}>0$. The observer-based event-triggered boundary control strategy consists of two components.
\begin{enumerate}
\item (The event-trigger) The set of event times $I=\{t_0,t_1,t_2,\ldots\}$ with $t_0=0$ forms an increasing sequence via the following rules: \begin{itemize}
\item if $\{t\in\mathbb{R}_{+}|t>t_{j}\wedge  d^{2}(t)> -\gamma m(t)\}=\emptyset$ then the set of the times of the events is $\{t_{0},\ldots,t_{j}\}.$
\item if $\{t\in\mathbb{R}_{+}|t>t_{j}\wedge d^{2}(t)> -\gamma m(t)\}\neq\emptyset$ then the next event time is given by:\begin{equation}\label{obetbc1}
t_{j+1}=\inf\{t\in\mathbb{R}_{+}|t>t_{j}\wedge d^{2}(t)>-\gamma m(t)\}
\end{equation} where $d(t)$ is given by\begin{equation}\label{obetbc2}
d(t)=\int_{0}^{1}k(y)\big(\hat{u}(y,t_{j})-\hat{u}(y,t)\big)dy,
\end{equation} for all $t\in[t_{j},t_{j+1})$ with $k(y)$ defined in \eqref{frk1}, and $m(t)$ satisfies the ODE 
\begin{equation}\label{obetbc3}\begin{split}
\dot{m}(t)=&-\eta m(t)+\rho d^{2}(t)-\beta_{1}\Vert\hat{w}[t]\Vert^{2}\\&-\beta_{2}\vert\hat{w}(1,t)\vert^{2}-\beta_{3}\vert\tilde{w}(1,t)\vert^{2},\end{split}
\end{equation}  for all $t\in(t_{j},t_{j+1})$ with $m(t_{0})=m(0)<0$ and $m(t_{j}^{-})=m(t_{j})=m(t_{j}^{+})$.  
\end{itemize}
\item (The control action) The output boundary feedback control law\begin{equation}\label{obetbc4}
U_{j}=\int_{0}^{1}k(y)\hat{u}(y,t_{j})dy,
\end{equation}for all $t\in[t_{j},t_{j+1}),j\in\mathbb{N}$.
\end{enumerate}
\end{dfn}

Proposition \ref{cor1} allows us to define the solution of the closed-loop system under the observer-based event-triggered boundary control \eqref{obetbc1}-\eqref{obetbc4} in the interval $[0,F)$, where $F=\sup(I)$.

\begin{lem}\label{lem1}Under the definition of the observer-based event-triggered boundary control \eqref{obetbc1}-\eqref{obetbc4}, it holds that $d^{2}(t)\leq-\gamma m(t)$ and $m(t)< 0,$ for $t\in [0,F)$ where $F=\lim_{j\rightarrow\infty} (t_j)$.
\end{lem}

\textit{Proof:} The proof is very similar to that of Lemma 1 in \cite{rathnayake2020observer} and hence omitted. 

\begin{lem}\label{lem2} For $d(t)$ given by \eqref{obetbc2}, it holds that\begin{equation}\label{ghm}
\dot{d}^{2}(t)\leq \rho_{1} d^{2}(t)+\alpha_{1}\Vert\hat{w}[t]\Vert^{2}+\alpha_{2}\vert\hat{w}(1,t)\vert^{2}+\alpha_{3}\vert\tilde{w}(1,t)\vert^{2},
\end{equation}
 for all $t\in (t_j,t_{j+1}),\in\mathbb{N}$, where  \begin{align}
\label{neqro}
\rho_{1}&=6\varepsilon^{2}k^{2}(1),\\
\begin{split}
\label{al1}\alpha_{1}&=3\tilde{L}^{2}\int_{0}^{1}\big(\varepsilon k''(y)+\varepsilon k(1)k(y)+\lambda k(y)\big)^{2}dy\\&\quad+6\big(\varepsilon qk(1)+\varepsilon k'(1)\big)^{2}\int_{0}^{1}L^{2}(1,y)dy,
\end{split}\\
\label{al2}
\alpha_{2}&=6\big(\varepsilon qk(1)+\varepsilon k'(1)\big)^{2},\\
\label{al3}
\alpha_{3}&=6\Big(\frac{\lambda k(1)}{2}+\int_{0}^{1}k(y)p_{1}(y)dy\Big)^{2}.
\end{align} with $k(y),\tilde{L},L$ given by \eqref{frk1}, \eqref{L1}, \eqref{Lsol}, respectively.\end{lem}

\textit{Proof:} The proof is very similar to that of Lemma 2 in \cite{rathnayake2020observer} and hence omitted.
\begin{thme}\label{iwm}Under the observer-based event-triggered boundary control in Definition \ref{def1}, with $\beta_{1},\beta_{2},\beta_{3}$ chosen as\begin{equation}\label{betas}
\beta_{1}=\frac{\alpha_{1}}{\gamma(1-\vartheta)},\hspace{5pt}\beta_{2}=\frac{\alpha_{2}}{\gamma(1-\vartheta)},\hspace{5pt}\beta_{3}=\frac{\alpha_{3}}{\gamma(1-\vartheta)},
\end{equation}where $\alpha_{1},\alpha_{2},\alpha_{3}$ given by \eqref{al1}-\eqref{al3} and $\vartheta\in(0,1)$,  there exists a minimal dwell-time $\tau>0$ between two triggering times, \textit{i.e.,} there exists a constant $\tau>0$ such that $t_{j+1}-t_{j}\geq\tau$, for all $j\in\mathbb{N}$, which is independent of the initial conditions and only depends on the system and control parameters.\end{thme}

\textit{Proof:} Let us define
\begin{equation}\label{fmdt}
\psi(t):=\frac{ d^{2}(t)+\gamma(1-\vartheta)m(t)}{-\gamma\vartheta m(t)}.
\end{equation}
Using Lemma \eqref{lem2}, a certain estimate for $\psi(t)$ can be obtained as in the proof of Theorem 1 in \cite{rathnayake2020observer}. Using this estimate, one can easily prove Theorem \ref{iwm}.
\begin{thme}\label{thm2}Let $\eta,\gamma>0$  be free design parameters, $\vartheta\in(0,1)$, and $g(x)$ and $r$ be given by \eqref{gt} and \eqref{rt}, respectively, while $\beta_{1},\beta_{2},\beta_3$ are chosen according to \eqref{betas}. Further, subject to Assumption \ref{ass1}, let us choose parameters  $B,\kappa_{1},\kappa_{2},\kappa_{3}>0$ such that
\begin{equation}\label{sapie1}
\begin{split}
B\bigg(\varepsilon\min\Big\{r,\frac{1}{2}\Big\}-\frac{\varepsilon}{2\kappa_{1}}-\frac{\lambda}{4\kappa_{2}}-\frac{\Vert g\Vert^{2}}{\kappa_{3}}\bigg)-2\beta_{1}-\beta_{2}>0,
\end{split}
\end{equation}and the design parameter
$\rho$ as
\begin{equation}\label{thmrho}
\rho=\frac{\varepsilon\kappa_{1} B}{2}.
\end{equation}
Then, the closed-loop system which consists of the plant  \eqref{ctpe1},\eqref{ctpe2},\eqref{mctpe3} and  the observer \eqref{ctoe1},\eqref{ctoe2},\eqref{mctoe3} with the  event-triggered boundary controller \eqref{obetbc1}-\eqref{obetbc4} has a unique solution and globally exponentially converges to zero, \text{i.e.,} $\Vert u[t]\Vert+\Vert\hat{u}[t]\Vert\rightarrow 0$ as $t\rightarrow \infty.$\end{thme}

\textit{Proof:} The proof is very similar to that of Theorem 2 in \cite{rathnayake2020observer} and hence omitted. Let us only provide the Lyapunov function used: $V=\frac{A}{2}\Vert\tilde{w}^{2}[t]\Vert^2+\frac{B}{2}\Vert\hat{w}^{2}[t]\Vert^2-m(t),
$ where $\hat{w}$ and $\tilde{w}$ are the systems described by \eqref{ettsm} and \eqref{etoet}, respectively, and $A>0$ is chosen such that $A> \frac{\lambda\kappa_2B+2\kappa_3 B+4\beta_3}{\varepsilon q}.$

\section{Numerical Simulations}
We consider a reaction-diffusion PDE with $\varepsilon=1;\lambda=10;q=5.1$ and the initial conditions $u[0]=10x^2(x-1)^{2}$ and $\hat{u}[0]=15x^2(x-1)^2+15x^3(x-1)^3$. For numerical simulations, both  plant and observer states are discretized with a uniform step size of $h=0.0062$ for the space variable. The discretization with respect to time was done using the implicit Euler scheme with step size $\Delta t=0.001\text{ }s$.

\begin{figure}
\centering
\includegraphics[scale=0.075]{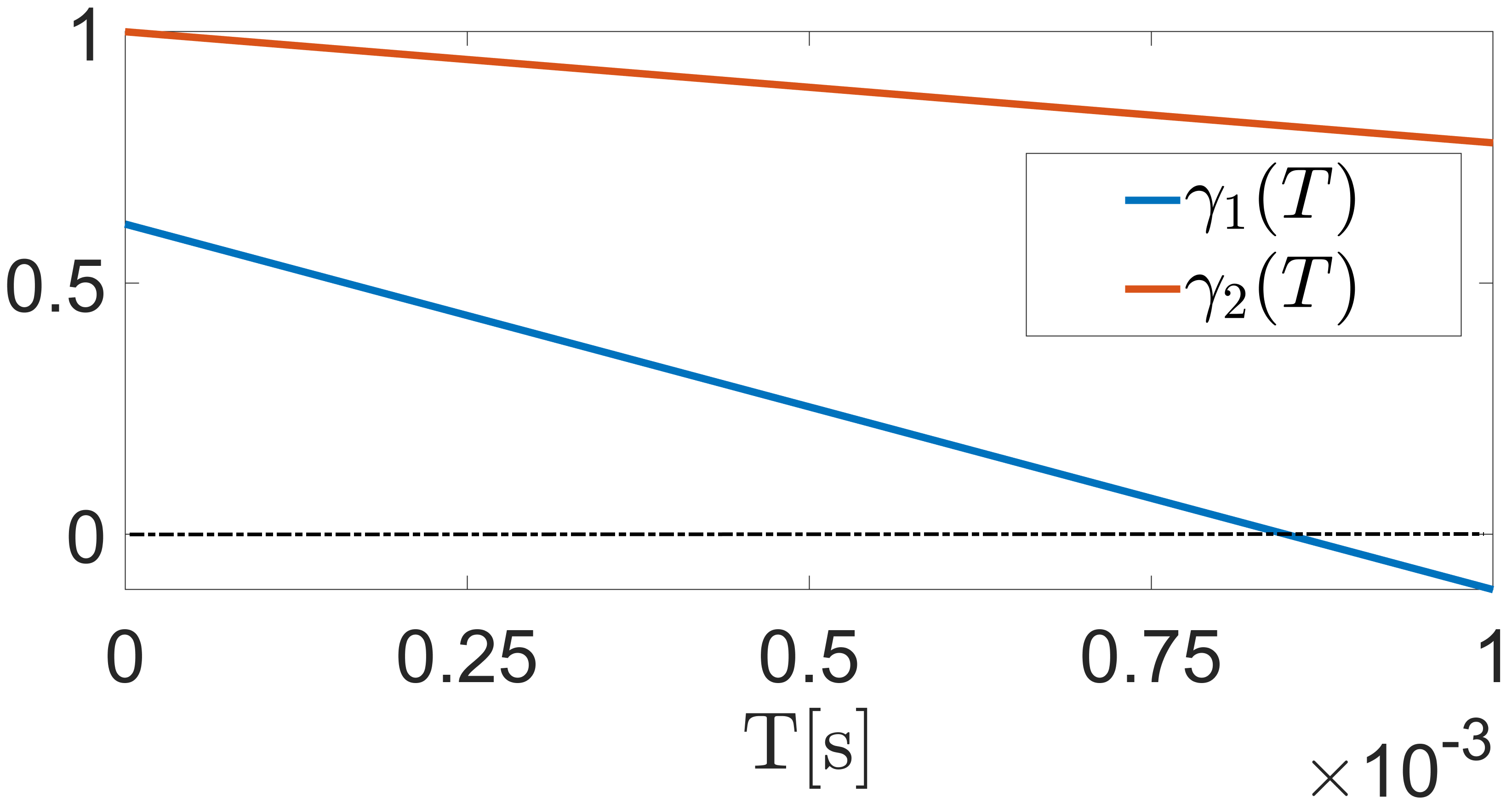}
\caption{The functions $\gamma_1(T)$ and $\gamma_2(T)$ given by \eqref{hn1} and \eqref{hn2} with $N=9$, $\sigma=0.0266$, and $C_1=0.5302$ in sampled-data control. A sampling diameter $T^{*}$ has to be selected such that $\gamma_1(T^{*})>0$ and $\gamma_2(T^{*})>0$ to guarantee exponential stability.}
\label{figT}
\end{figure}

First, we look at the sampling diameter for the sampled-data control implementation of the considered reaction-diffusion system. The functions $\gamma_1(T)$ and $\gamma_2(T)$ with respect to $T$, given by \eqref{hn1} and \eqref{hn2}, respectively, are shown in Fig. \ref{figT}. We have used $N=9$, $\sigma=0.0266$, and $C_1=0.5302$ in $\gamma_1(T)$ and $\gamma_2(T)$. As a sampling diameter $T^{*}$ should be chosen such  that $\gamma_1(T^{*})>0$ and $\gamma_2(T^{*})>0$ (see \eqref{set1}), the maximum possible sampling diameter is around $8\times 10^{-4}\text{ }s$, which is very small.

The parameters for the event-trigger mechanism are chosen as follows: $m(0)=-0.5,\gamma=10^{5},\eta=1\text{ or }100$ and  $\vartheta=0.1$. We can compute using \eqref{al1}-\eqref{al3} that $\alpha_{1}=1.3511\times 10^{3};\alpha_2= 1.9642\times 10^2;\alpha_3=1.1956\times 10^4$. Therefore, from \eqref{betas}, we can obtain $\beta_{1}=0.015;\beta_{2}= 0.0022;\beta_{3}=  0.1328$. Finding that $\Vert g\Vert^{2}=  3.0042\times 10^4$, let us choose $\kappa_1=11;\kappa_2=10^4;\kappa_3=10^8$ and $B=  0.6440$ to satisfy \eqref{sapie1}. Then, from \eqref{thmrho}, we can obtain $\rho=3.54$.

\begin{figure}
\centering
\subfloat[]{\includegraphics[scale=.065]{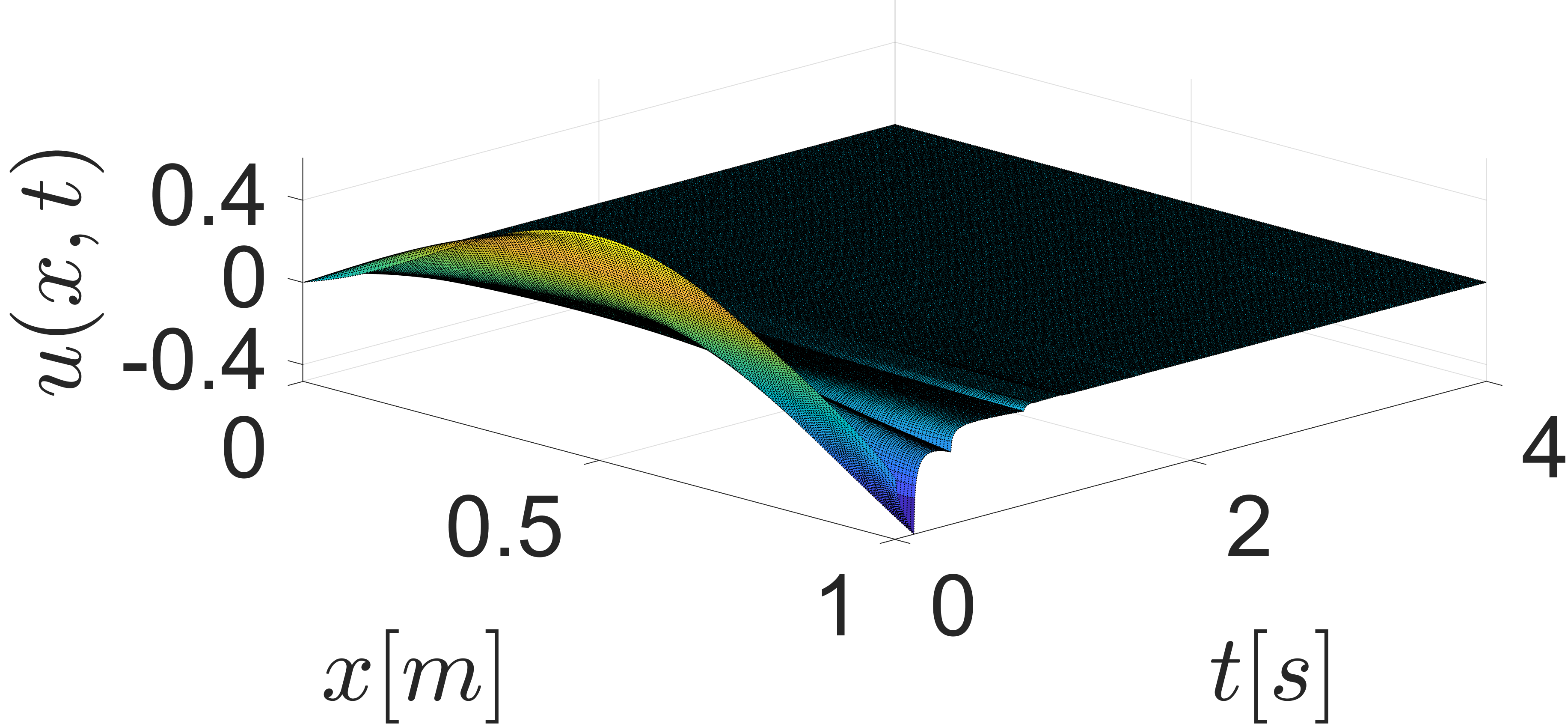}}\\ \subfloat[]{\includegraphics[scale=.065]{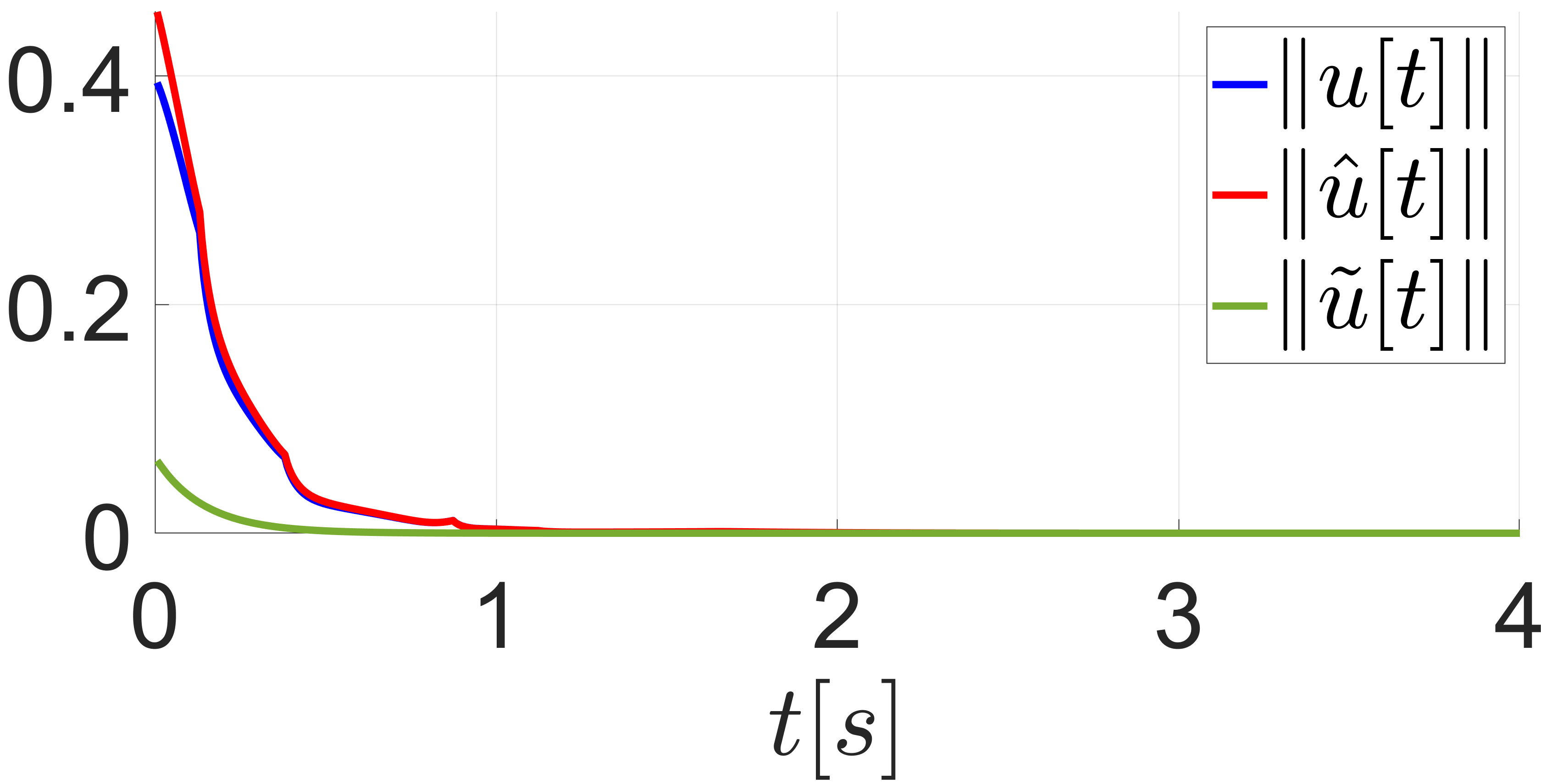}}
\caption{Results for the event-triggered closed-loop system with $\varepsilon=1,\lambda=10,q=5.1,m(0)=-0.5,\eta=1$, $u[0]=10x^2(x-1)^{2}$ and $\hat{u}[0]=15x^2(x-1)^2+15x^3(x-1)^3$ (a) $u(x,t)$. (b) $\Vert u[t]\Vert\text{ and }\Vert\tilde{u}[t]\Vert.$ }
\label{3dplt}
\end{figure}

\begin{figure}
\centering
\includegraphics[scale=0.075]{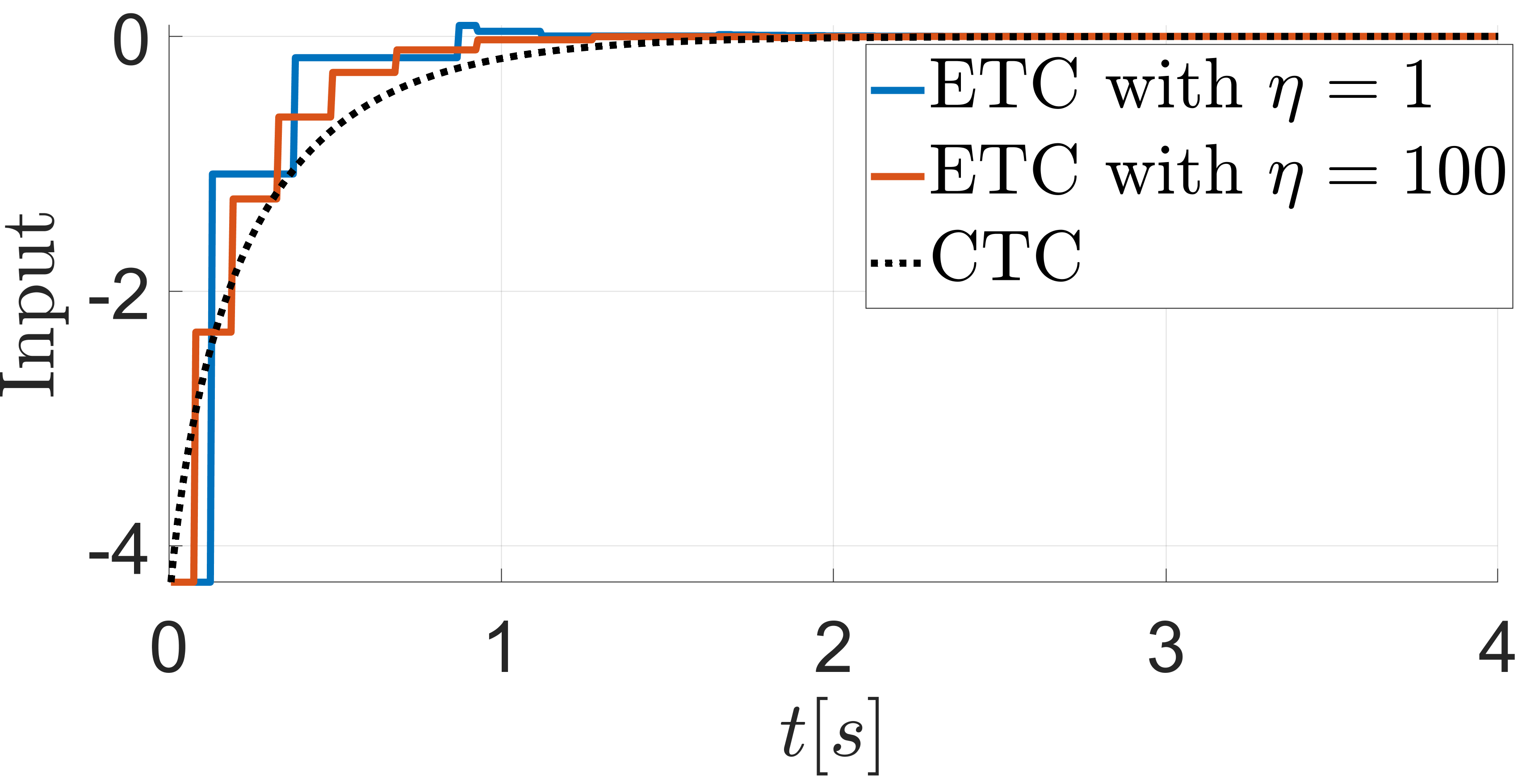}
\caption{Comparison of ETC (event-triggered control) input for different $\eta$: $\eta=1\text{ and }\eta=100$, for the same system considered in Fig. \ref{3dplt}. The corresponding continuous-time control (CTC) input is also provided for reference.} 
\label{figs}
\end{figure}

Fig. \ref{3dplt}(a) shows the response of the pant under event-triggered control with $\eta=1$  and Fig. \ref{3dplt}(b) shows the resulting $\Vert u[t]\Vert, \Vert\hat{u}[t]\Vert$, and $\Vert\tilde{u}[t]\Vert$. The evolution of the control inputs when $\eta=1$ and $\eta=100$ is presented in Fig. \ref{figs} along with the corresponding continuous-time control input. It can be observed that $\eta=100$ yields faster sampling than $\eta=1$. 

\section{Conclusion}

This paper has proposed sampled-data and event-triggered boundary control strategies for a class of reaction-diffusion systems with collocated sensing and Robin actuation. For sampled-data control, it has been shown that the continuous-time output-feedback backstepping boundary control applied in a sample-and-hold fashion guarantees global closed-loop exponential stability, provided that the sampling period is sufficiently small. Robustness with respect to perturbations of the sampling schedule is guaranteed. For the event-triggered implementation of the continuous-time control, a dynamic triggering condition has been used in order to determine when the control value needs to be updated.  Under the event-triggered control approach, the existence of a minimal-dwell time between two updates independent of initial conditions has been proved. Further, the global exponential convergence of the closed-loop system to the equilibrium has been established. 

\bibliographystyle{elsarticle-num}
\bibliography{main}

\begin{thebibliography}{10}
\expandafter\ifx\csname url\endcsname\relax
  \def\url#1{\texttt{#1}}\fi
\expandafter\ifx\csname urlprefix\endcsname\relax\def\urlprefix{URL }\fi
\expandafter\ifx\csname href\endcsname\relax
  \def\href#1#2{#2} \def\path#1{#1}\fi

\bibitem{hespanha2007survey}
J.~P. Hespanha, P.~Naghshtabrizi, Y.~Xu, A survey of recent results in
  networked control systems, Proceedings of the IEEE 95~(1) (2007) 138--162.

\bibitem{fridman2004robust}
E.~Fridman, A.~Seuret, J.-P. Richard, Robust sampled-data stabilization of
  linear systems: an input delay approach, Automatica 40~(8) (2004) 1441--1446.

\bibitem{karafyllis2009global}
I.~Karafyllis, C.~Kravaris, Global stability results for systems under
  sampled-data control, International Journal of Robust and Nonlinear Control:
  IFAC-Affiliated Journal 19~(10) (2009) 1105--1128.

\bibitem{nesic2009explicit}
D.~Nesic, A.~R. Teel, D.~Carnevale, Explicit computation of the sampling period
  in emulation of controllers for nonlinear sampled-data systems, IEEE
  Transactions on Automatic Control 54~(3) (2009) 619--624.

\bibitem{karafyllis2011nonlinear}
I.~Karafyllis, M.~Krstic, Nonlinear stabilization under sampled and delayed
  measurements, and with inputs subject to delay and zero-order hold, IEEE
  Transactions on Automatic Control 57~(5) (2011) 1141--1154.

\bibitem{karafyllis2012global}
I.~Karafyllis, M.~Krstic, Global stabilization of feedforward systems under
  perturbations in sampling schedule, SIAM Journal on Control and Optimization
  50~(3) (2012) 1389--1412.

\bibitem{pepe2016stability}
P.~Pepe, On stability preservation under sampling and approximation of
  feedbacks for retarded systems, SIAM Journal on Control and Optimization
  54~(4) (2016) 1895--1918.

\bibitem{tabuada2007event}
P.~Tabuada, Event-triggered real-time scheduling of stabilizing control tasks,
  IEEE Transactions on Automatic Control 52~(9) (2007) 1680--1685.

\bibitem{heemels2012introduction}
W.~Heemels, K.~H. Johansson, P.~Tabuada, An introduction to event-triggered and
  self-triggered control, in: 2012 IEEE 51st IEEE Conference on Decision and
  Control (CDC), IEEE, 2012, pp. 3270--3285.

\bibitem{6069816}
M.~C.~F. {Donkers}, W.~P. M.~H. {Heemels}, Output-based event-triggered control
  with guaranteed $\mathcal{L}_{\infty}$-gain and improved and decentralized
  event-triggering, IEEE Transactions on Automatic Control 57~(6) (2012)
  1362--1376.

\bibitem{heemels2012periodic}
W.~H. Heemels, M.~Donkers, A.~R. Teel, Periodic event-triggered control for
  linear systems, IEEE Transactions on Automatic Control 58~(4) (2013)
  847--861.

\bibitem{marchand2012general}
N.~Marchand, S.~Durand, J.~F.~G. Castellanos, A general formula for event-based
  stabilization of nonlinear systems, IEEE Transactions on Automatic Control
  58~(5) (2013) 1332--1337.

\bibitem{tallapragada2013event}
P.~Tallapragada, N.~Chopra, On event triggered tracking for nonlinear systems,
  IEEE Transactions on Automatic Control 58~(9) (2013) 2343--2348.

\bibitem{postoyan2014framework}
R.~Postoyan, P.~Tabuada, D.~Ne{\v{s}}i{\'c}, A.~Anta, A framework for the
  event-triggered stabilization of nonlinear systems, IEEE Transactions on
  Automatic Control 60~(4) (2015) 982--996.

\bibitem{girard2014dynamic}
A.~Girard, Dynamic triggering mechanisms for event-triggered control, IEEE
  Transactions on Automatic Control 60~(7) (2015) 1992--1997.

\bibitem{logemann2003stability}
H.~Logemann, R.~Rebarber, S.~Townley, Stability of infinite-dimensional
  sampled-data systems, Transactions of the American Mathematical Society
  355~(8) (2003) 3301--3328.

\bibitem{logemann2005generalized}
H.~Logemann, R.~Rebarber, S.~Townley, Generalized sampled-data stabilization of
  well-posed linear infinite-dimensional systems, SIAM journal on Control and
  Optimization 44~(4) (2005) 1345--1369.

\bibitem{rebarber2006robustness}
R.~Rebarber, S.~Townley, Robustness with respect to sampling for stabilization
  of {R}iesz spectral systems, IEEE Transactions on Automatic Control 51~(9)
  (2006) 1519--1522.

\bibitem{fridman2012robust}
E.~Fridman, A.~Blighovsky, Robust sampled-data control of a class of semilinear
  parabolic systems, Automatica 48~(5) (2012) 826--836.

\bibitem{selivanov2017sampled}
A.~Selivanov, E.~Fridman, Sampled-data relay control of diffusion {{PDE}}s,
  Automatica 82 (2017) 59--68.

\bibitem{karafyllis2017sampled}
I.~Karafyllis, M.~Krstic, Sampled-data boundary feedback control of {1-D}
  linear transport {PDE}s with non-local terms, Systems \& Control Letters 107
  (2017) 68--75.

\bibitem{karafyllis2018sampled}
I.~Karafyllis, M.~Krstic, Sampled-data boundary feedback control of {1-D}
  parabolic {{PDE}}s, Automatica 87 (2018) 226--237.

\bibitem{kang2018distributed}
W.~Kang, E.~Fridman, Distributed sampled-data control of
  {K}uramoto--{S}ivashinsky equation, Automatica 95 (2018) 514--524.

\bibitem{wang2019mixed}
J.-W. Wang, J.-M. Wang, Mixed ${H}_2/ {H}_\infty$ sampled-data output feedback
  control design for a semi-linear parabolic {PDE} in the sense of spatial
  ${L}_\infty$ norm, Automatica 103 (2019) 282--293.

\bibitem{katz2021delayed}
R.~Katz, E.~Fridman, Delayed finite-dimensional observer-based control of {1-D}
  parabolic {{PDE}}s, Automatica 123 (2021) 109364.

\bibitem{selivanov2016distributed}
A.~Selivanov, E.~Fridman, Distributed event-triggered control of diffusion
  semilinear {PDE}s, Automatica 68 (2016) 344--351.

\bibitem{espitia2016event}
N.~Espitia, A.~Girard, N.~Marchand, C.~Prieur, Event-based control of linear
  hyperbolic systems of conservation laws, Automatica 70 (2016) 275--287.

\bibitem{espitia2017event}
N.~Espitia, A.~Girard, N.~Marchand, C.~Prieur, Event-based boundary control of
  a linear 2x2 hyperbolic system via backstepping approach, IEEE Transactions
  on Automatic Control 63~(8) (2018) 2686--2693.

\bibitem{wang2019observer}
J.-W. Wang, Observer-based boundary control of semi-linear parabolic {{PDE}}s
  with non-collocated distributed event-triggered observation, Journal of the
  Franklin Institute 356~(17) (2019) 10405--10420.

\bibitem{espitia2020observer}
N.~Espitia, Observer-based event-triggered boundary control of a linear
  2$\times$ 2 hyperbolic systems, Systems \& Control Letters 138 (2020) 104668.

\bibitem{espitia2021event}
N.~Espitia, I.~Karafyllis, M.~Krstic, Event-triggered boundary control of
  constant-parameter reaction--diffusion {PDEs}: A small-gain approach,
  Automatica 128 (2021) 109562.

\bibitem{diagne2020event}
M.~Diagne, I.~Karafyllis, Event-triggered control of a continuum model of
  highly re-entrant manufacturing system, arXiv preprint arXiv:2004.01916
  (2020).

\bibitem{katz2020boundary}
R.~Katz, E.~Fridman, A.~Selivanov, Boundary delayed observer-controller design
  for reaction-diffusion systems, IEEE Transactions on Automatic Control
  (2020).

\bibitem{karafyllis2019input}
I.~Karafyllis, M.~Krstic, Input-to-state stability for {{PDE}}s, Springer,
  2019.

\bibitem{rathnayake2020observer}
B.~Rathnayake, M.~Diagne, N.~Espitia, I.~Karafyllis, Observer-based
  event-triggered boundary control of a class of reaction-diffusion {PDE}s,
  IEEE Transactions on Automatic Control (2021).

\bibitem{chen2017backstepping}
J.~Chen, B.~Zhuang, Y.~Chen, B.~Cui, Backstepping-based boundary feedback
  control for a fractional reaction diffusion system with mixed or robin
  boundary conditions, IET Control Theory \& Applications 11~(17) (2017)
  2964--2976.

\end{thebibliography}

\end{document}